\documentclass{WS-IJMPE}
\usepackage{graphicx}
\usepackage{dcolumn}
\usepackage{bm}
\usepackage{amsfonts}

\begin{document}


\markboth{A.V. Nazarenko}{Probability of creation of extra dimensions...}

\title{PROBABILITY OF CREATION OF EXTRA DIMENSIONS IN NUCLEAR COLLISIONS}

\author{\footnotesize A.V. NAZARENKO}

\address{Bogolyubov Institute for Theoretical Physics,\\
14-b, Metrologichna Str., Kiev 03680, Ukraine\\
nazarenko@bitp.kiev.ua}

\maketitle

\begin{history}
\received{(received date)}
\revised{(revised date)}
\end{history}

\begin{abstract}
The minisuperspace model in $3+d$ spatial dimensions with the matter,
described by the bag model, is considered with the aim to estimate the
probability of creation of the compactified extra dimensions in nuclear
collisions. The amplitude of transition from 3- to $3+d$-dimensional
space has been calculated both in the case of completely confined matter,
when contribution of radiation is ignored, and in the case of radiation
domination, when the bag constant is negligible. It turns out that the number
of additional dimensions is limited in the first regime, while this is infinite
in the second one. It is shown that the probability of creation of extra
dimensions is finite in the both regimes.
\end{abstract}


\section{Introduction}

As it is expected, the Large Hadron Collider (LHC) will be able to detect
quantum gravitational effects. Theoretically, the high-energy particle beam at the
LHC can cool by boiling off gravitons into the extra space-like dimensions.\cite{ANP}
More exotic gravitational objects, such as micro black holes, are assumed to be also
produced at the TeV energies (see Refs.\cite{Land,BHHS,St} and references therein).
In fact the LHC can explore these additional dimensions through apparent breaking
the energy conservation, as well as the appearance and disappearance of gravitons
from extra dimensions as envisioned by the superstring theory.

Expectations to observe strongly gravitational effects at the TeV energies mainly
follow from Ref.\cite{ADD}, where it is supposed that the electroweak scale
$m_{\rm EW}\sim10^3$~GeV is the only fundamental short distance scale in nature.
If the gravity is unmodified down to the Planck length, one might assume that the
fundamental mass $M_{\rm f}$ in $(3+d)+1$-dimensional theory is of $m_{\rm EW}$ order
at the some value of $d$ (the number of the compactified extra dimensions within
the framework of Kaluza-Klein scenario). As is pointed out in Ref.\cite{ADD}, to
satisfy this requirement, parameter $d$ should be equal to 2. At $d=0$, the usual
Plank mass, $M_{\rm Pl}=\sqrt{\hbar c/(16\pi G)}\approx1.72\cdot10^{18}$~GeV, is
restored. Moreover, such an approach allows one to determine the fundamental scale
$L$ of extra dimensions:
\begin{equation}\label{Mass}
L^d\sim\frac{M^2_{\rm Pl}}{M^{2+d}_{\rm f}}.
\end{equation}

Because the inverse square law of gravitational attraction has only been verified at
distances greater than $1$~mm (see Ref.\cite{LP} and references therein), it is now
suggested that the huge disparity in
interaction strengths is caused by the ``leaking'' of gravitational field into the
extra dimensions which can be as large as $1$~mm without violating known
laboratory, astrophysical, or cosmological data. In this picture, the particles and
forces are confined to a 3-brane embedded into multi-dimensional universe, the only force
in the higher-dimensional bulk is gravity.

Since there is no evidence of the presence of extra dimensions in low-energy
(classical and quantum) processes, one can assume that an appearance of them is
induced by the highly dense state of energy and matter like a thermalized quark-gluon
plasma, predicted by quantum chromodynamics (QCD). Thus, the purpose of this paper is to
determine the probability of creation of additional dimensions in ultrarelativistic
heavy-ion collisions.

The mechanism of creation of extra dimensions can be divided in two steps:
i) generating the gravitational field in our three spatial dimensions by the matter
source; ii) transferring the part of energy of this field component into the extra
dimensions, where new (extra) modes of gravitational field arise. Such a
mechanism is similar to the transfer of the ``shaking'' (deformation) between
slices within the elasticity theory. Note that this analogy looks straightforward
because there is mathematical relation between general relativity and elasticity
theory.\cite{Sah,Kok,Pad}

Here we develop the phenomenological model including the matter source,
described by the bag model, and the gravitational field, represented by two
scale factors reflecting two physically different groups of dimensions,
the topology of which is fixed. The coupling constant (gravitational constant
in $(3+d)+1$-dimensional space-time) in this model is defined from Eq.~(\ref{Mass}).

Although we will apply the methods and achievements of quantum cosmology, but
there are several differences between it and our model. Here we assume that
the 4-dimensional Universe already exists while the quantum cosmology is
dealing with the creation and evolution of Universe from ``nothing''.

\section{Model Formulation}

Let us focus on formulation of the model containing two interacting participants:
gravitational field and the dense matter being in the quark-gluon phase. As usual,
we will assume that the homogeneous and isotropic matter lives in 3-dimensional
space while the gravitational field propagates in $3+d$ dimensions. Of course,
the one time-like evolution parameter is also presented in the model and is
common for matter and field. We then can imagine that the dense matter, produced
in nuclear collisions, plays the role of source of gravitational field in three
dimensions, whose excitation induces an emergence of the gravitational field
modes in extra dimensions, the number of which we would like to estimate.

Let the matter be concentrated in the fireball of radius $r_{\rm f}
\approx 7$~fm, and all of three-dimensional phenomena under our consideration are
carried out in the bulk of this region. The thermodynamic properties (pressure $P$
and energy density $\varepsilon$) of the quark-gluon phase are described within the
framework of the bag model,\cite{GSS} taking into account only light quarks (and gluons),
\begin{equation}\label{matter}
P=\frac{\pi^2}{90}gT^4-B,\quad
\varepsilon=\frac{\pi^2}{30}gT^4+B,
\end{equation}
where $T$ is the temperature, $g=37$ is degeneracy factor, $B^{1/4}=200$~MeV is
the bag constant.

To evaluate the probability of creation of extra spatial dimensions, we should
obviously to appeal to the quantum picture which provides the appropriate
framework to reach this goal. More precisely, we should deal with the case of
quantum gravity and/or cosmology in order to find quantum states of (micro) universe
and to calculate an amplitude of transition between 3- and $3+d$-dimensional spaces.

Here we will not account for the topology and geometry variations of extra space,
which may emerge. Even making use of the gravitational path integral, the sum over all
topologies is not computable, since there is no algorithm for deciding whether two
$3+d$-dimensional manifolds have the same topology. Moreover, there is the problem
(even in dimension two) that there exist many more geometries of complicated topology
than there exist of simple topology. It makes impossible to define the theory
non-perturbatively in an unambiguous and physically satisfactory manner. In higher
dimensions these problems are unmanagable. For these reasons we are forced
to consider the model with the fixed topology of (extra) space, what allows us to
apply canonical quantization of gravitational field.

Note that canonical quantization is useful for investigations of the system dynamics,
although it requires to impose boundary conditions and to solve the problem of the
operator ordering. These problems are general for quantum cosmology (minisuperspace)
models, based on Wheeler--DeWitt equation, and have conceptual meaning in the context
of the Universe evolution understanding.

Coming back to the problem of fixing topology, let the extra dimensions be
compactified and described by periodic variables $\phi_n\in(0,2\pi L)$, where $L$ is the
fundamental length of the higher dimensional theory, $n=\overline{1,d}$. Then extra
$d$-dimensional manifold $\Sigma_d$ is compact and isomorphic to
\begin{equation}\label{top}
\underbrace{\Bbb{S}\times\Bbb{S}\ldots\times\Bbb{S}}_{d}.
\end{equation}

Recall that in the original Kaluza-Klein models, space-time was taken to be, at least
locally, a product manifold of the form $\Bbb{R}_4\times\Sigma_d$, where $\Sigma_d$
is some compact ``internal space'', in the first instance a circle. The effective
space-time $\Bbb{R}_4$ must thus be identified with the quotient of the higher
dimensional space-time with respect to $\Sigma_d$.\footnote{Sometimes, authors identify
our space-time with a submanifold rather than a quotient.}

For the sake of simplicity, we can exploit a common scale factor, say $b$, for
additional dimensions. If $b=0$, these dimensions are absent, and we concern with the
theory in four-dimensional space-time. Such an approximation does not permit us to
investigate the geometry variations for the given topology. Nevertheless, it may be
effective if the sizes of dimensions are roughly of the same order.

Similarly, it is possible to introduce a common scale factor for our three dimensions.
In the other words, it means that the 3-space is supposed to be conformally flat and
isotropic.

At $x^2+y^2+z^2<r^2_{\rm f}$, where the matter is highly dense, the potentials of
homogeneous gravitational field, $a$ and $b$, are introduced as follows
\begin{equation}
ds^2=-dt^2+a^2(t)(dx^2+dy^2+dz^2)+b^2(t)\sum\limits_{n=1}^d d\phi^2_n,
\end{equation}
where we use the gauge-fixing condition ($N=1$, where $N$ is the lapse function)
resulting in the formalism with ``Hubble time''.

On the other hand, the metric becomes asymptotically Minkowskian and $d=0$ (and $b=0$) at
$x^2+y^2+z^2\gg r^2_{\rm f}$.

The quantum states of this model are described by wave function $\langle a,b|3+d\rangle$
dependent on $d$. Therefore the amplitude of transition $"3"\to"3+d"$  at the fixed
$b$, playing role of parameter, is defined as
\begin{equation}\label{ampl}
\left|\int d\mu(a)\langle 3|a,0\rangle\langle a,b|3+d\rangle\right|,
\end{equation}
where $\mu(a)$ is the integration measure.
 
Having got the normalized wave function $\langle a,b|3+d\rangle$, we can also find more
probable values of $b$ at the fixed $a$ and $d$ from the probability density profile,
\begin{equation}
|\langle a,b|3+d\rangle|^2.
\end{equation}

Another possibility to describe extra dimensions consists in the finding of the
wave function $\langle 3+d|a,b_1,b_2,\ldots,b_d\rangle$ dependent on $1+d$
scale factors. Such an approach could be used to study the different configurations
(geometries) of $3+d$-dimensional space. However, we limit ourselves here by the
simple case depicted before.

\section{Hamiltonian Dynamics}

In this Section, starting from an action of relativistic system
``gravitational field+ideal fluid'', we reformulate the dynamics in the terms
of canonical variables needed for further canonical quantization.

As mentioned in previous Section, we assume that the matter is located in the
region of 3-dimensional space at $x^2+y^2+z^2<r^2_{\rm f}$. Therefore there
is the following energy conservation in three spatial dimensions:
\begin{equation}\label{law}
\dot\varepsilon+3\frac{\dot a}{a}(\varepsilon+P)=0,
\end{equation}
where the dot means derivative with respect to $t$.

Since the thermodynamic properties of the matter are described by Eqs.~(\ref{matter}),
resulting in equation of state,
\begin{equation}
P=\frac{1}{3}\varepsilon-\frac{4}{3}B,
\end{equation}
integrating Eq.~(\ref{law}), one obtains that the dependence of the energy density
$\varepsilon$ on the scale factor is
\begin{equation}
\varepsilon=B+\frac{\cal E}{a^4},
\end{equation}
where ${\cal E}=\pi^2gT^4/30$, and the scale factor at $t=t_0$ is assumed to be $a_0=1$.

Thus, the action functional of the (quark-gluon) ideal fluid in the rest frame of the
matter is of the form:
\begin{eqnarray}
S_{\rm M}&=&-\int\varepsilon a^3dtd^3x
\nonumber\\
&=&-V_{\rm f}\int\left(Ba^3+\frac{\cal E}{a}\right)dt,
\end{eqnarray}
where $V_{\rm f}=4\pi r^3_{\rm f}/3$ is the fireball volume.

Note that the successive derivation of $S_{\rm M}$ can be done, for instance, by
means of formalism elaborated in Ref.\cite{Hama}

To write down the action functional for the gravitational field, we have to find
the scalar curvature of $(3+d)+1$-dimensional space-time. The term containing the
four-dimensional curvature is often missing from the action. However, in general,
this term seems to be essential since it is generated as a quantum correction to
the matter action. Note that this quantum correction typically involves an infinite
number of terms of higher order in curvature. Nevertheless, these terms are
completely suppressed in our consideration and the action functional is
formulated in accordance with Einstein.

The non-vanishing Cristoffel symbols of second kind are
\begin{eqnarray}
&& \Gamma^0_{xx}=\Gamma^0_{yy}=\Gamma^0_{zz}=a\dot a,\qquad
\Gamma^0_{\phi_n \phi_n}=b\dot b,
\nonumber\\
&&
\Gamma^x_{0x}=\Gamma^y_{0y}=\Gamma^z_{0z}=\frac{\dot a}{a},\qquad
\Gamma^{\phi_n}_{0\phi_n}=\frac{\dot b}{b},
\end{eqnarray}
and the Ricci scalar is presented in the following form:
\begin{equation}
R[a,b]=G[a,b]+\frac{2}{a^3b^d}\frac{d^2(a^3b^d)}{dt^2},
\end{equation}
where one can ignore the last term in r.h.s., resulting in total derivative with respect to
evolution parameter. We have introduced the notation: 
\begin{equation}
G[a,b]=-6\frac{\dot a^2}{a^2}-6d\frac{\dot a}{a}\frac{\dot b}{b}-d(d-1)\frac{\dot b^2}{b^2}.
\end{equation}

As was discussed above, the gravitational field is supposed to be homogeneous and 
non-trivial inside three-dimensional fireball while the metric in the space beyond
fireball tends asymptotically to the usual Minkowskian one (without additional dimensions).
Therefore it is possible to limit the region of integration in the action functional
by the volume filled by the quark-gluon matter. Such an approximation means that we are
neglecting by the surface contribution, where the gradient of pressure, gravitational
potentials and curvature, in principal, can be large enough. This approximation is valid,
if the profiles of matter distribution and scale factors are preserved. It is easy to
understand that the matter distribution changes are related with longitudinal hydrodynamic
waves, propagating with the sound speed $c_{\rm s}\equiv\sqrt{dP/d\varepsilon}\approx0.57c$.
Then the characteristic time of these changes can be evaluated as 
$\tau_{\rm c}=r_{\rm f}/c_{\rm s}\approx12.3$~fm/c and is approximately equal to the
life-time of fireball. Therefore we conclude the time of process must be limited,
$\tau_{\rm c}\gg\Delta t$.

Thus the Lagrangian function of the system, which we consider, is written as
\begin{equation}
L=V_{\rm f}\left[M^2_{\rm Pl}G[a,b]a^3b^d-U(a)\right],
\end{equation}
where we have applied the relation $M^{2+d}_{\rm f}=M^2_{\rm Pl}/\int d^d\phi_n$
and introduced the matter potential $U(a)=Ba^3+{\cal E}/a$.

Canonical momenta conjugated to scale factors $a$ and $b$ respectively are
\begin{eqnarray}
p_a&\equiv&\frac{\partial L}{\partial \dot a}=-6M^2_{\rm Pl}V_{\rm f}a^2b^d
\left(2\frac{\dot a}{a}+d\frac{\dot b}{b}\right),\\
p_b&\equiv&\frac{\partial L}{\partial \dot b}=-2dM^2_{\rm Pl}V_{\rm f}a^3b^{d-1}
\left((d-1)\frac{\dot b}{b}+3\frac{\dot a}{a}\right).
\end{eqnarray}

Resolving these relations with respect to velocities $\dot a$, $\dot b$, one finds that
\begin{eqnarray}
\dot a&=&\frac{1}{2(2+d)M^2_{\rm Pl}V_{\rm f}ab^d}\left(\frac{d-1}{3}p_a-\frac{b}{a}p_b\right),\\
\dot b&=&\frac{1}{d(d+2)M^2_{\rm Pl}V_{\rm f}a^2b^{d-1}}\left(\frac{b}{a}p_b-\frac{d}{2}p_a\right).
\end{eqnarray}

Canonical Hamiltonian of the system is
\begin{eqnarray}
H&\equiv&\dot ap_a+\dot bp_b-L\nonumber\\
&=&\frac{d(d-1)a^2p^2_a-6dabp_ap_b+6b^2p^2_b}{12d(d+2)M^2_{\rm Pl}V_{\rm f}a^3b^d}
+V_{\rm f}U(a).
\end{eqnarray}
Introducing the operators instead of classical variables, we will use this function for
formulating Wheeler-DeWitt equation in the next Section.

\section{Quantization. Probability of Creation of Extra Dimensions}

Before quantization is performed, let us make some modifications.

Due to general covariance, the system at the classical level should be described by
Hamiltonian constraint $H=0$, which can be re-written as
\begin{equation}
\frac{12}{d(d+2)}\left(bp_b-\frac{d}{2}ap_a\right)^2
=a^2\left[p^2_a-\varkappa^2ab^dU(a)\right],
\end{equation}
where we have introduced the coefficient $\varkappa^2\equiv24M^2_{\rm Pl}
V^2_{\rm f}/(\hbar c)^6\approx2.5\cdot10^{36}$~MeV${}^{-4}$.

Since there is no correct limit for Hamiltonian constraint at $d\to0$, we perform
a canonical transformation. It is convenient to introduce new variables
$\xi_d=ab^{d/2}$, $v_d=b^d$ (dependent on $d$) instead of scale factors $a$, $b$.
The conjugate momenta are related as
\begin{eqnarray}
&&p_\xi=p_ab^{-d/2},\quad
p_v=\frac{1}{db^d}\left(bp_b-\frac{d}{2}ap_a\right),
\\
&&p_a=\sqrt{v_d}p_\xi,\quad
p_b=dv_d^{1-1/d}\left(p_v+\frac{\xi_d}{2v_d}p_\xi\right).
\end{eqnarray}

In the terms of new canonical variables, we obtain
\begin{equation}
\frac{12d}{d+2}v_d^2p^2_v
=\xi^2_d\left[p^2_\xi-\varkappa^2\frac{\xi_d}{\sqrt{v_d}}U\left(\frac{\xi_d}{\sqrt{v_d}}\right)\right].
\end{equation}

At $d=0$, this equation is reduced to the following one:
\begin{equation}
p^2_\xi-\varkappa^2\frac{\xi_0}{\sqrt{v_0}}U\left(\frac{\xi_0}{\sqrt{v_0}}\right)=0,
\end{equation}
where $\xi_0=a$, $v_0=1$, $p_\xi=p_a$ as must be.

Making use of the scheme of canonical quantization, one has to replace momentum variables
with differential operators: 
\begin{equation}
p_{\xi}\to-i\frac{\partial}{\partial\xi_d},\quad
p_v\to-i\frac{\partial}{\partial v_d}.
\end{equation}

The Wheeler-DeWitt equation with the normal ordering of operators (which is usual
for quantum field theory) is written as follows
\begin{equation}
\left[\xi^2_d\frac{\partial^2}{\partial\xi^2_d}-
\frac{12d}{d+2}v_d^2\frac{\partial^2}{\partial v^2_d} +\varkappa^2\frac{\xi^3_d}{\sqrt{v_d}}U\left(\frac{\xi_d}{\sqrt{v_d}}\right)\right]\Psi_d=0.
\end{equation}

At $d=0$, it is reduced to the ordinary differential equation:
\begin{equation}
\left[\frac{\partial^2}{\partial a^2} +\varkappa^2aU(a)\right]\Psi_0=0.
\end{equation}

Note that this connection makes sense due to the form of (space-time interval and)
Hamiltonian where parameter $d$ can be of fractal (non-integer) dimension. Moreover,
if it is possible to change $d$ continuously, the topology (\ref{top}) of extra
dimensions becomes also fractal. However, we will not discuss here aspects of
the fractal topology construction.

In some cases of $U(x)$, these equations can be solved exactly. However, since
$\varkappa^2U(x)$ is quite large in our consideration, the quasiclassical WKB
approximation is applicable enough.

Note also that, we are not imposing boundary conditions in our minisuperspace model.
Hence such a problem formulation can be treated as pseudo-Euclidean analogy of the
Hartle--Hawking ``no boundary'' proposal.

In the next subsections we will investigate the particular regimes of the bag model
equation of state.

\subsection{The Case of Confined Matter}

Let $B\gg{\cal E}$ and, therefore, $U(a)\approx Ba^3$. In this case (of cold matter),
the matter pressure is negative and is approximately equal to $-B$. The dynamics of
gravitational field is strongly damped in this regime. Just decreasing $B$, a
sufficient role of the field can be expected. However, such a (confined) state of
matter, predicted by QCD, should be investigated for better understanding of
phenomena, which emerge under unification of strong and gravitational interactions.

Now let us describe briefly the general scheme used in further calculations.
Limiting ourselves by a first quantum correction, the wave functions are found
in the form:
\begin{equation}
\Psi\sim{\rm e}^{iS/\hbar},\quad
S\approx S_0+\frac{\hbar}{i}S_q,
\end{equation}
where we write down explicitly Plank constant $\hbar$; $S_0$ is the classical
action, which satisfies the Hamilton--Jacobi equation:
\begin{equation}
\frac{12d}{d+2}v_d^2\left(\frac{\partial S_0}{\partial v_d}\right)^2
=\xi^2_d\left[\left(\frac{\partial S_0}{\partial\xi_d}\right)^2
-\varkappa^2\frac{\xi_d}{\sqrt{v_d}}U\left(\frac{\xi_d}{\sqrt{v_d}}\right)\right].
\end{equation}

The first quantum correction $S_q$ is found from equation:
\begin{equation}
2\xi^2_d\frac{\partial S_0}{\partial\xi_d}\frac{\partial S_q}{\partial\xi_d}
+\xi^2_d\frac{\partial^2S_0}{\partial\xi^2_d}=\frac{12d}{d+2}
\left(2v^2_d\frac{\partial S_0}{\partial v_d}\frac{\partial S_q}{\partial v_d}
+v^2_d\frac{\partial^2S_0}{\partial v^2_d}\right).
\end{equation}

Substituting $U(a)$, the wave functions in WKB approximation are
\begin{equation}
\Psi_d\sim\frac{v_d}{\xi_d}\exp{\left(i\varkappa\sqrt{\frac{B}{3}\frac{2+d}{6-d}}\frac{\xi^3_d}{v_d}\right)},
\end{equation}
\begin{equation}
\Psi_0\sim\frac{1}{a}\exp{\left(i\varkappa\frac{\sqrt{B}}{3}a^3\right)}
=\frac{\sqrt{v_d}}{\xi_d}\exp{\left(i\varkappa\frac{\sqrt{B}}{3}\frac{\xi^3_d}{v^{3/2}_d}\right)}.
\end{equation}

Looking at $\Psi_d$, we see that the number of extra dimensions is limited by $d=6$.
At $d>6$ the wave function becomes real and exponentially decreasing. Since
$\varkappa\sqrt{B}\approx6.3\cdot10^{23}$, $\Psi_{d>6}$ vanishes.
On the other hand, the wave function
\begin{equation}
\Psi^*_d\sim\frac{v_d}{\xi_d}\exp{\left(-i\varkappa\sqrt{\frac{B}{3}\frac{2+d}{6-d}}\frac{\xi^3_d}{v_d}\right)},
\end{equation}
becomes exponentially increasing at $d>6$ and, therefore, this state is
unphysical. It confirms that the number of extra dimensions in this model does not exceed
six.

Using the wave functions found, let us consider the transition amplitude from 3- to
$3+d$-dimensional space at the fixed $v_d$. We assume that the all compact dimensions are
opened i.e., $v_d=1$. In this case, the wave functions are described by a common
expression:
\begin{equation}
\langle\xi_d|3+d\rangle=\frac{\sqrt{\Xi}}{\xi_d}\exp{\left(i\varkappa\sqrt{\frac{B}{3}\frac{2+d}{6-d}}\xi^3_d\right)}.
\end{equation}
This wave function is normalized with respect to the integration measure:
\begin{equation}
\int\limits_\Xi^\infty|\langle\xi_d|3+d\rangle|^2d\xi_d=1,
\end{equation}
where the lower limit is chosen to be $\Xi>0$. If $\Xi=1$, one has exactly the
scale factor of Minkowski space-time. It means that the integration over the
region $\xi_d\in(\Xi,\infty)$ accounts for the expanding geometry only. The
process of collapsing (when $\xi_d\in(0,\Xi)$) should be ignored in the system
with negative pressure in order to eliminate unphysical situation. Mathematically,
the integration in the range $\xi_d\in(0,\Xi)$ brings a divergent term.

The transition amplitude is defined as
\begin{equation}
{\cal T}(d)\equiv\left|\int\limits_\Xi^\infty \langle 3|\xi_d\rangle\langle\xi_d|3+d\rangle d\xi_d\right|.
\end{equation}

The result of computations is presented as
\begin{equation}\label{prob1}
{\cal T}(d)=\left|1+izU\left(1,\frac{5}{3},-iz\right)\right|,
\end{equation}
where $U(a,b,z)$ is the Kummer function,
\begin{equation}
z=\varkappa\frac{\sqrt{B}}{3}\Xi^3\left(\sqrt{3\frac{2+d}{6-d}}-1\right).
\end{equation}

At $z\to\infty$ and $d>0$, the asymptotic expression for the transition
amplitude is ${\cal T}\approx1/(3z)$. At $d\to0$ ($z\to0$), we
have to use Eq.~(\ref{prob1}) and obtain ${\cal T}(0)=1$.

Since the number of extra dimensions is limited, we can construct the continuous
probability density as
\begin{equation}
{\cal P}(d)={\cal T}^2(d)\left/\int\limits_0^6{\cal T}^2(x)dx\right.,
\end{equation}
which depends on the model parameter $\mu\equiv\varkappa\sqrt{B}\Xi^3/3$.
Here we will not explore the question: what happens if we put $d<0$.

The behavior of ${\cal P}(d)$ is demonstrated in Fig.~1. We may
immediately conclude that the probability of creation of few extra dimensions
is {\it finite}!
\begin{figure}[htbp]
\begin{picture}(80,110)
\put(70,-50){\includegraphics[width=8cm,angle=0]{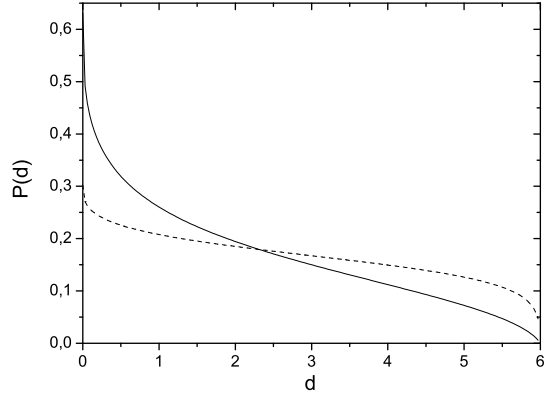}}
\end{picture}
\vspace*{10mm}
\caption{Probability density of creation of extra dimensions at $\mu=0.11$
(solid line), 0.01 (dashed line).}
\end{figure}

Now, making use of the probability density ${\cal P}(d)$, it is interesting to
investigate the average number of additional dimensions:
\begin{equation}
\langle d\rangle=\int\limits_0^6 x{\cal P}(x)dx.
\end{equation}
This is function of parameter $\mu$. Thus, we get that $\langle d\rangle\to3$,
when $\mu\to0$, and $\langle d\rangle\to0$ at $\mu\to\infty$. In order to
reproduce $\langle d\rangle=2$, what is predicted by ADD paradigm, one has
$\mu\approx0.11$. In principal, we should treat $\Lambda_{\rm eff}\equiv B\Xi^6$
as the vacuum energy density of QCD in the present regime. To realize
$\langle d\rangle=2$, we get that $\Lambda_{\rm eff}\approx4.36\cdot10^{-38}$~MeV${}^4$.
Since the cosmological constant is often interpreted as the vacuum energy,
one can conclude that this quantity $\Lambda=3.956\cdot10^{-35}$~MeV${}^4$ 
(following Refs.\cite{WMAP1,WMAP2}) is larger than $\Lambda_{\rm eff}$ obtained.

\subsection{The Case of Radiation Domination}

Now let us assume ${\cal E}\gg B$ and $U(a)\approx{\cal E}/a$. It means that
the temperature of the quark-gluon gas is much higher than $B^{1/4}=200$~MeV.
This case is an interesting one because the energy density of matter (and the
matter temperature) can be changed due to gravitational effects like an origin
of additional dimensions, while the bag constant is supposed to be unchanged.

Since the energy density ${\cal E}$ is the function of the temperature $T$,
which may be changed, let us introduce quasi-momentum $k=\varkappa\sqrt{\cal E}$
as the free parameter and the characteristic quasi-momentum $K\approx2.2\cdot10^{23}$
for $T=200$~MeV. In principal, $k$ should be bigger than $K$ in the regime
under consideration.

The Wheeler--DeWitt equations without boundary conditions for the system,
where radiation dominates, are solved exactly and give us the wave functions:
\begin{eqnarray}
\Psi_0&\sim&\exp{(ika)}=\exp{\left(ik\frac{\xi_d}{\sqrt{v_d}}\right)},\\
\Psi_d&\sim&v^\lambda_d\sqrt{\xi_d}
H^{(1)}_{\frac{1}{2}\sqrt{1+\frac{48d}{d+2}\lambda(\lambda-1)}}\left(k\xi_d\right),
\label{ex}
\end{eqnarray}
where $H^{(1)}_\nu$ is the Hankel function; $\lambda$ is arbitrary number such that
$\Re\lambda>0$.

Note that $\Psi_d$ is infinite at $\xi_d=0$, if $\lambda(\lambda-1)>0$. Indeed,
the further analysis is proving that $\lambda(\lambda-1)<0$ and $\lambda=1/2-i\mu$,
$\mu\in\Bbb{R}$. In the last case, the index of the Hankel function is purely
imaginary number, resulting in vanishing the wave function at $\xi_d=0$.

Remembering that $k>10^{23}$, let us consider an asymptotic expression (up to
constant phase) for $\Psi_d$:
\begin{equation}\label{psi}
\Psi_d\sim v^\lambda_d\exp{(ik\xi_d)}\left(1+i\lambda(\lambda-1)\frac{6d}{d+2}\frac{1}{k\xi_d}
+O\left(\frac{1}{(k\xi_d)^2)}\right)\right),
\end{equation}
where we will omit the term of order $O\left(1/(k^2\xi_d^2)\right)$.

On the other hand, the WKB approach leads to the following classical action and
the first quantum correction:
\begin{eqnarray}
S_0&=&\sqrt{c+\xi^2_dk^2}-\sqrt{c}\ln{\left(2\frac{c+\sqrt{c}\sqrt{c+\xi^2_dk^2}}{\xi_d}\right)}
-\frac{\sqrt{c}}{2}\sqrt{\frac{d+2}{3d}}\ln{v_d}+C_1,
\\
S_q&=&\frac{1}{2}\ln{\frac{\xi_d}{\sqrt{c+\xi^2_dk^2}}}
-\sqrt{\frac{3d}{d+2}}\ln{\left(2\frac{c+\sqrt{c}\sqrt{c+\xi^2_dk^2}}{\xi_d}\right)}+C_2,
\end{eqnarray}
where $c$, $C_1$, $C_2$ are arbitrary constants.

Using these formulas, we try to analyze the consistency of the approximate solutions and
to connect the constants.

First of all, let us expand $S_0$ and $S_q$ in the same manner like $\Psi_d$:
\begin{eqnarray}
S_0&=&C_1-\sqrt{c}\ln{(2k\sqrt{c})}-\frac{\sqrt{c}}{2}\sqrt{\frac{d+2}{3d}}\ln{v_d}+\xi_dk
-\frac{c}{2\xi_dk}+O\left(\frac{c^2}{k^2\xi_d^2}\right),
\\
S_q&=&C_2-\frac{1}{2}\ln{k}-\sqrt{\frac{3d}{d+2}}\ln{(2k\sqrt{c})}-\sqrt{\frac{3d}{d+2}}\frac{\sqrt{c}}{\xi_dk}
+O\left(\frac{1}{k^2\xi_d^2}\right).
\end{eqnarray}

Comparing $\exp{(iS_0+S_q)}$ with $\Psi_d$ from Eq.~(\ref{psi}), it is convenient to take
\begin{equation}
C_1=\sqrt{c}\ln{(2k\sqrt{c})},\quad
C_2=\frac{1}{2}\ln{k}+\sqrt{\frac{3d}{d+2}}\ln{(2k\sqrt{c})},\quad
c=\frac{3d}{d+2}s^2,
\end{equation}
where $s$ is regarded as a new parameter instead of $c$.

Comparing this expression with Eq.~(\ref{psi}) again, we derive that
$s=2i\lambda$ and $\lambda(\lambda-1)=-(s^2-2is)/4$. In order to be
$s^2-2is\in\Bbb{R}$ (the Hamiltonian operator has to be (quasi)Hermitian),
one finds parametrization, $s=2\mu+i$, resulting in $s^2-2is=1+4\mu^2$,
where $\mu\in\Bbb{R}$. 

We interpret the case of $\mu=0$ as the ground state of the system. Without
loosing generality, we will focus on this situation. It is clear that
the excited states are less probable than the case at a hand.

Now let us discuss the normalization condition for the wave functions. We will
assume again that $v_d=1$. Then the scale factor $\xi_d$ can run from 0 to $\infty$
in this case because there is no problem with the negative pressure and collapse.
Because of $\Bbb{Z}_2$ symmetry of the space-time interval (with respect to the sign
inversion of scale factors), the range of $\xi_d$ can be further extended to
$(-\infty,+\infty)$.

Note that the asymptotic expression for $\Psi_d$ has the pole at $\xi_d=0$. This
property is important for normalization procedure. To determine the integration
contour near $\xi_d=0$, we appeal to the exact solution (\ref{ex}): the symmetry of
$|\Psi_d|^2$ profile in the ranges $\xi_d\in(-\infty,0)$ and $\xi_d\in(0,+\infty)$
is preserved only if $\xi_d\to\xi_d+i0^+$.

Using well-known formula for Dirac $\delta$-function:
\begin{equation}
\frac{1}{2\pi}\int\limits_{-\infty}^\infty{\rm e}^{ikx}dx=\delta(k),
\end{equation}
the wave function $|3,k\rangle=\exp{(ik\xi_d)}/\sqrt{2\pi}$ at $d=0$ is normalized as
\begin{equation}
\langle3,q|3,k\rangle=\delta(k-q).
\end{equation}
This relation reflects simply the energy conservation in three dimensions.

The normalization condition for the wave function at $d>0$,
\begin{equation}
|3+d,k\rangle=\frac{1}{\sqrt{2\pi}}\exp{(ik\xi_d)}\left(1-i\frac{3d}{2(d+2)}\frac{1}{k\xi_d}\right),
\end{equation}
looks like
\begin{eqnarray}
\langle3+d,q|3+d,k\rangle&\approx&\frac{1}{2\pi}\int\limits_{-\infty}^\infty d\xi_d{\rm e}^{i\xi_d(k-q)}
\left[1-i\frac{3d}{2(d+2)}\left(\frac{1}{k}-\frac{1}{q}\right)\frac{1}{\xi_d+i0^+}\right]
\\
&=&\delta(k-q)-\frac{3d}{2(d+2)}\frac{q-k}{kq}\theta(q-k).
\end{eqnarray}
We see that the energy density of matter can be varied in the presence of extra dimensions!

Note that, taking term $O(1/(k^2\xi_d^2))$ into account in
$|3+d,k\rangle$, one gets term of $O((k-q)^2/(k^2q^2))$ order
after integration. Therefore, our approximation is valid for the small
energy deviation $|k-q|/k$. The $\theta$-function is defined such that
$\theta(0)=1/2$.

Having got all necessary ingredients, we are able to calculate the amplitude of
transition from 3- to $3+d$-dimensional space. This quantity is
\begin{equation}\label{ta}
\langle3,q|3+d,k\rangle=\delta(k-q)-\frac{3d}{2(d+2)}\frac{1}{k}\theta(q-k).
\end{equation}
The Heaviside $\theta$-function in the last term corresponds to the cooling of the matter.
It is understandable that this process is related with the transferring of energy from the
matter to the extra dimensions.

Moreover, there is another effect consisting in the changes of the system volume.
If $v_d\not=1$, we have to replace $q$ with $q/\sqrt{v_d}$ in the last formula.

The probability of creation of extra dimensions in the case of radiation domination is
defined as
\begin{equation}
{\cal W}=\int\limits_0^\infty|\langle3,q|3+d,k\rangle|^2dk.
\end{equation}
As usual,\cite{BLP} we should make the following replacement:
\begin{equation}
\delta^2(k-q)\to\delta(k-q)\frac{1}{2L}\int\limits_{-L}^L{\rm e}^{i\xi_d(k-q)}d\xi_d=\delta(k-q).
\end{equation}

Limiting by the term of order $1/q$, one finds
\begin{equation}
{\cal W}=1-\frac{3d}{2(d+2)}\frac{1}{q}.
\end{equation}
This formula explains why the high energy (or temperature $T$) is needed to create
extra dimensions. However, the model predicts the infinite number of additional
dimensions in the contrast with the case of the confined matter. Since
\begin{equation}
{\cal W}_{d\to\infty}=1-\frac{3}{2q},
\end{equation}
it is impossible to find the probability density in our approximation.

\section{Conclusions}

In this paper the minisuperspace model in $3+d$ spatial dimensions with the matter,
described by the bag model, has been considered with the aim to estimate the
probability of the origin and the number of the compactified extra dimensions in nuclear
collisions. The gravitational constant in $(3+d)+1$ space-time is introduced on
the base of ADD paradigm,\cite{ADD} and the topology of extra dimensions is
supposed to be toroidal in accordance with Kaluza--Klein scenario.

For the sake of simplicity, the amplitude of transition from 3- to
$3+d$-dimensional space has been calculated in two regimes of the bag model:
i) in the case of completely confined matter (what is equal to the model with
the cosmological constant); ii) in the case of radiation domination, when the bag
constant is negligible.

In the first case, the number of additional dimensions is limited by $d=6$,
that is agreed with superstring theory prediction.

It is shown in the second case that the energy density (the temperature) of the matter
is changed in the presence of extra dimensions, the number of which is not limited
in approximation used.

It turns out that it is possible to evaluate the mean value of extra dimensions in
the first case only and to obtain $\langle d\rangle=2$ in agreement with ADD
predictions.

It is important to emphasize that the probability of creation of extra dimensions is
finite in the both regimes. The probability for intermediate regimes is not
computed here and should be investigated.

\section*{Acknowledgments}
\noindent
I am greatly indebted to Yu.~Kulinich (Lviv National University) for many fruitful
discussions and for some constructive comments in the filed of cosmology.



\end{document}